\documentstyle[preprint,aps]{revtex}
\begin{document}
\draft
\title{Gauss Law Constraints in Chern-Simons Theory From BRST Quantization }
\author{M. Chaichian${}^{a,b}$, W. F. Chen${}^{b}$\renewcommand{\thefootnote}
{\dagger}\footnote{\small ICSC-World Laboratory, Switzerland}
 and Z.Y. Zhu${}^{c,d}$}
\address{${}^a$ High Energy Physics Laboratory, Department of Physics\\
${}^b$ Research Institute for High Energy Physics, 
University of Helsinki\\
P.O. Box 9 (Siltavuorenpenger 20 C), FIN-00014 Helsinki, Finland\\
${}^c$ CCAST(World Laboratory), P.O. Box 8730, Beijing 100080, China\\
${}^d$ Institute of Theoretical Physics, Academia Sinica,
P. O. Box 2735, Beijing, 100080, China}

\maketitle

\begin{abstract}
The physical state condition in the BRST quantization of Chern-Simons field
theory is used to derive Gauss law constraints in the presence of Wilson
loops, which play an important role in explicitly establishing the connection
of Chern-Simons field theory with 2-dimensional conformal field theory.
\end{abstract}


\vspace{3ex}

When we discuss knot invariants in terms of Chern-Simons theory$^{\cite
{wit,gua,van,roza,horn,kg}}$ and the relationship between Chern-Simons field
theory and conformal field theory$^{\cite{agg,ms}}$, an important relation
is Gauss law constraint in presence of Wilson line, which was first given 
in ref.${\cite{wit}}$. This relation plays an important role in proving that
states of Chern-Simons theory satisfy the 
Knizhnik-Zamolodchikov equation$^{\cite{bn,lr,emss,gins,kz}}$. 
In this letter we intend to derive the Gauss
law constraints from BRST quantization of Chern-Simons field theory. We
think this investigation is significant since in some sense BRST
quantization formulation is defined better than a formal manipulation without
gauge fixing$^{\cite{kugo}}$. We will show that when Wilson lines exist the
physical state condition in BRST quantization will lead to Gauss law
constraints with source terms just as those given in ref. ${\cite
{wit}}$. The procedure we will adopt is similar to the one in ref.${\cite{mar}%
}$, where the equivalence between Dirac's first-class constraints and BRST
treatment for Yang-Mills theory is formally proved.

Let us first write down the BRST quantization of Chern-Simons theory. The
action of Chern-Simons field theory takes the following form 
\begin{equation}
S_{\text{CS}}=\displaystyle\frac k{4\pi }
{\int }_{M^3}\text{Tr}\left[ A{\wedge }dA+\frac 23A{%
\wedge }A{\wedge }A\right]~ ,
\end{equation}
where $A=A_{\mu}dx^{\mu}=A_\mu ^aT^adx^\mu $ 
with $T^a$ being the generators in some
representation of gauge group $G$. The parameter $k$ must be chosen to be an
integer in order to make the theory gauge invariant under large gauge
transformations. Without loss of generality, we choose $G=SU(N)$ and
the normalization $\text{Tr}(T^aT^b)=\displaystyle\frac{1}{2}$. Choosing
the Lorentz gauge ${\partial }^\mu A_\mu ^a=0$ and performing BRST gauge
fixing, we obtain the following effective action 
\begin{eqnarray}
\displaystyle S_{\text{eff}} &=&{\int}d^3x~{\cal L}_{\text{eff}}
=S_{\text{CS}}+{\int }\text{Tr}{\delta }[\bar{c}({\partial }_\mu
A^\mu +B)]  \nonumber \\
&=&\displaystyle {\int }d^3x\left\{\frac k{16\pi }{\epsilon }^{\mu \nu \rho
}\left[ A_\mu ^a({\partial }_\nu A_\rho ^a-{\partial }_\rho A_\nu ^a)+i\frac 
23f^{abc}A_\mu ^aA_\nu ^bA_\rho ^c\right] \right. \nonumber \\
&&\displaystyle \left.-\frac{ik}{8\pi }A^{{\mu }a}{\partial }_\mu B^a+\frac{ik}{%
8\pi }(B^a)^2-\frac 12{\partial }_\mu \bar{c}^aD^\mu c^a\right\}~,
\end{eqnarray}
where $B=B^aT^a$ is the auxiliary field and $A_{\mu}=A_{\mu}^aT^a,
c=c^aT^a, \bar{c}=\bar{c}^aT^a$. The BRST transformations of the 
fields are as follows 
\begin{equation}
\begin{array}{l}
{\delta }A_\mu ^a=D_\mu c^a,~{\delta }B^a=0~, \\[2mm] 
\displaystyle {\delta }c^a=-\frac 12f^{abc}c^bc^c,
~{\delta }\bar{c}^a=\frac{ik}{4\pi }B^a~.
\end{array}
\end{equation}
These transformations are nilpotent, i.e., ${\delta }^2=0$. Now, obviously
the classical configuration space is enlarged by the introduction of new
fields---ghost fields $c^a$, anti-ghost fields $\bar{c}^a$ and multiplier
fields $B^a$. The canonically conjugate momenta can be well defined by $%
\displaystyle {\Pi }_\Phi =\frac{{\partial }{\cal L}}{{\partial }{\dot{\Phi}}%
}$, with ${\Phi }={\{}A_1,B,c,\bar{c}{\}}$: 
\begin{eqnarray}
\displaystyle {\Pi }_{A_1}^a &=&\frac k{8\pi }A_2^a,~{\Pi }_B^a=-i\frac k{%
8\pi }A_0^a~,  \nonumber \\
\displaystyle {\Pi }_{\bar{c}}^a &=&-\frac 12D^0c^a,
~{\Pi }_c^a=\frac 12\dot{\bar{c}}^a~.
\end{eqnarray}
These fields and their canonically conjugate momenta satisfy the 
Poisson brackets
(for bosonic fields) or anti-brackets (for fermionic fields): 
\begin{eqnarray}
{\{}{\Pi }_\Phi ^i({\bf x},t)~,~{\Phi }_j({\bf y},t){\}}_{{\pm}\text{PB}}&=&-i{%
\delta }_j^i{\delta }^{(2)}({\bf x}-{\bf y})~,  \nonumber \\
{\{}{\Pi }_\Phi ^i({\bf x},t)~,~{\Pi }_\Phi ^j({\bf y},t){\}}_{{\pm}\text{PB}}
&=&{\{}{\Phi }^i({\bf x},t)~,~{\Phi }^j({\bf y},t){\}}_{{\pm}\text{PB}}=0~.
\end{eqnarray}
The BRST charge can be obtained by the Noether theorem 
\begin{eqnarray}
\displaystyle Q &=&{\int }d^2x\left[ \frac k{8\pi }{\epsilon }%
^{ij}D_ic^aA_j^a-\frac 14f^{abc}\dot{\bar{c}}^ac^bc^c-\frac{ik}{8\pi }%
B^aD^0c^a\right]  \nonumber \\
\displaystyle &=&{\int }d^2x\left[ -\frac k{8\pi }c^aF_{12}^a
-\frac{1}{2}f^{abc}{\Pi }_c^ac^bc^c
+\frac{ik}{4\pi }B^a{\Pi }_{\bar{c}}^a\right]~ .
\label{brstcharge}
\end{eqnarray}
It is easy to show that 
\begin{equation}
\displaystyle {\delta }{\Phi}={\{}Q~,~{\Phi }{\}}_{{\pm}\text{PB}},
~{\{}Q~,~Q{\}}_{{\pm}\text{PB}}=Q^2=0~.
\end{equation}
When we perform quantization, the classical observables are replaced by
operators, and (anti-) Poisson brackets by (anti-) commutative Lie brackets.
With the present polarization choice, the Hilbert space are composed of
square integrable functionals of ${\Phi }$. The quantum BRST charge operator 
$\hat{Q}$ is nilpotent 
\begin{equation}
\displaystyle \frac 12{\{}\hat{Q},\hat{Q}{\}}=\hat{Q}^2=0~,
\end{equation}
where a hat ``$\hat{~}$" denotes an operator. It is well known that the 
state space here possesses indefinite metric.
According to the general principle of BRST quantization, physical states
satisfy the so-called ``BRST-closed'' condition 
\begin{equation}
\hat{Q}|\text{phys}{\rangle }=0~.  \label{physical}
\end{equation}
Notice that above condition (\ref{physical}) determines a physical state up
to ``BRST-exact states'', i.e. 
\begin{equation}
|\text{phys}{\rangle }{\sim }|\text{phys}{\rangle }+|\chi {\rangle },~|\chi {%
\rangle }=\hat{Q}|\text{any states}{\rangle }~.
\end{equation}
Obviously these states $|\chi {\rangle }$ are normal to all physical states
including themselves, 
\begin{equation}
{\langle }\chi |\text{phys}{\rangle }={\langle }{\chi }_1|{\chi }_2{\rangle }%
=0~.
\end{equation}
Thus they are zero norm states and make no contribution to the
physical observables. Now we define the physical operator $\hat{\Phi}$ to be
an operator that generates a physical state from vacuum. It is easy to show
that the physical operator $\hat{\Phi}$ must satisfy the condition 
\begin{equation}
\lbrack \hat{\Phi},\hat{Q}]_{\pm }=f[\hat{\Phi}]\hat{Q}
\end{equation}
due to Eq.(\ref{physical}). Furthermore, the operators can be divided into
two classes. According to ref.${\cite{mar}}$, we call them as the A-type and
the B-type. An A-type operator transforms a physical state into another one 
\begin{equation}
\hat{A}|\text{phys}{\rangle }=|\text{phys}{\rangle }^{\prime }~.
\end{equation}
A B-type operator transforms a physical state into a BRST exact state and
has the form 
\begin{equation}
\hat{B}=[*,\hat{Q}]_{\pm }~,  \label{btype}
\end{equation}
where $*$ represents some operator. Eq.(\ref{btype}) implies that a B-type
operator can be regarded as the generator of a kind of gauge transformation
since it does not affect physical observables. The (anti-)commutators of
B-type operators with an arbitrary physical operator $\hat{\Phi}$ have the
form 
\begin{equation}
\lbrack \hat{B}_i,\hat{\Phi}]_{\pm }=g[\hat{\Phi}]_{ij}\hat{B}_j~,
\end{equation}
which means that B-type operators form an ideal in the operator algebra$^{%
\cite{mar}}$ 
\begin{equation}
\begin{array}{l}
\lbrack \hat{A},\;\hat{A}]_{\pm }\subset \{\hat{A}~\&~\hat{B}\}~, \\[2mm] 
\lbrack \hat{A},\;\hat{B}]_{\pm }\subset \{\hat{B}\},\;[\hat{B},\;\hat{B}%
]_{\pm }\subset \{\hat{B}\}~.
\end{array}
\end{equation}
The product of an arbitrary operator $\hat{K}$ (physical or nonphysical)
with a B-type operator can also be regarded as the generator of gauge
transformations due to the fact that
\begin{equation}
\lbrack (\hat{K}\hat{B})_i,\hat{\Phi}]_{\pm }=h[\hat{\Phi}]_{ij}(\hat{K}\hat{%
B})_j~.
\end{equation}
In addition, $\hat{K}\hat{B}$ operators also form a closed algebra 
\begin{equation}
\lbrack (\hat{K}\hat{B})_i,\hat{K}\hat{B})_j]_{\pm }=U_{ij}^k(\hat{K}\hat{B}%
)_k~.
\end{equation}
Notice that the $\hat{K}\hat{B}$ operator transforms a physical state out of
the genuine physical state space$^{\cite{mar}}$. We can see in the following
that the properties of $\hat{B}$ or $\hat{K}\hat{B}$ operators play a crucial
role in our derivation.

From the BRST charge given in Eq.(\ref{brstcharge}) we can show that 
\begin{equation}
\begin{array}{l}
\displaystyle \hat{B}_1^a\equiv [\hat{Q},\hat{\Pi}_c^a]=-\frac k{8\pi }\hat{F%
}_{12}^a-\frac 12f^{abc}\hat{\Pi}_c^b\hat{c}^c~, \\[2mm] 
\displaystyle \hat{B}_2^a\equiv [\hat{Q},\hat{\Pi}_B^a]=\frac k{4\pi }\hat{%
\Pi}_c^a~, \\[2mm] 
\displaystyle \hat{B}_3^a\equiv [\hat{Q},{\partial }^\mu \hat{A}_\mu
^a]=M_{ab}\hat{c}^b~,\label{decomp}
\end{array}
\end{equation}
where $\displaystyle M_{ab}=\frac k{8\pi }[\hat{F}_{12}^a,{\partial }^\mu 
\hat{A}_\mu ^b]$. Furthermore, 
we know that the matrix $(M_{ab})$ is non-singular from
the fact that $\hat{F}_{12}$ and ${\partial }^\mu \hat{A}_\mu $ constitute a
pair of second-class constraints$^{\cite{dirac,hnne}}$. Note that in
deriving the Eqs.(\ref{decomp}) we have used the $B$-field 
equation of motion (on-shell
condition). The non-singularity of $M$ ensures that ghost field operators
can be written as $\hat{c}^a=(M^{-1})^{ab}\hat{B}_3^b$ and belong to the 
B-type.  Hence they are indeed the generators of gauge transformations.

From Eqs.(\ref{brstcharge}), (\ref{decomp}) and the above arguments, one can see
that the three terms composed of BRST charge $\hat{Q}$ are all gauge
transformation generators. However the second and third terms are $\hat{B}$-
or $\hat{K}\hat{B}$-type operators. Thus when $\hat{Q}$ acts 
on physical states, the
second and the third terms transform the physical state to non-physical
state. Since there exists no coupling between the nonphysical gauge
transformation generators $\hat{c}^a$ and the physical ones, after the
action of BRST charge, the transformed state $|~{\rangle }$ can be written as 
\begin{equation}
|~{\rangle }=|\text{non-phys}{\rangle \oplus }|\text{phys}{\rangle }~.
\end{equation}
So the physical state condition $\hat{Q}|\text{phys}{\rangle }=0$ reduces to 
\begin{equation}
\displaystyle \hat{F}_{12}^a|phys{\rangle }=0~,
\label{gausslaw}
\end{equation}
when no Wilson loop exists. Now we turn to the case in the presence of
Wilson loops. Let us take the manifold $M=\Sigma {\times }R$ as in ref.${\cite
{wit}}$, where $R$ is the time variable space and $\Sigma $ is the spatial
surface. The physical state at some time $t$ in 
the presence of a Wilson loop can be
represented by a punctured surface $\Sigma $, the puncture points being
produced by the intersections of the surface $\Sigma $ with the links where
the Wilson loop operators are defined.  This has been 
given in ref.${\cite{gmm1}}$
\begin{equation}
\displaystyle
|\text{phys}{\rangle }={\Pi }_{n=1}^N
exp\left[i{\int }_{P_n~({\Gamma }_n)}^{Q_n}
{\sum }_{i=1,2}\hat{A}_i^{(n)}({\bf x})dx^i\right]|0{\rangle }~,
\end{equation}
where $n$ denotes the $n$th component of links, ${\Gamma }_n$ is the
projection on $\Sigma $ of links located in the three dimensional
space-time region less than 
time $t$ and $P_n$, $Q_n$ are the endpoints of ${\Gamma }_n$. 
In the polarization chosen above, 
the state functional can be
written down explicitly in a path integral form$^{\cite{gmm1}}$ 
\begin{equation}
\begin{array}{l}
\displaystyle {\Psi }_{\text{phys}}[\Phi]
={\langle}\hat{\Phi}|\text{phys}{\rangle}\\[2mm]
\displaystyle =\left({\int }{\cal D}{\Xi}^{\prime}~\left\{ exp\left[ i{\int 
}_{-{\infty }}^tdt^{\prime}{\int }_\Sigma d^2x{\cal L}_{\text{eff}}
-\frac{k}{2\pi}{\int }_\Sigma d^2x
{\sum }_{i=1}^2{A^{\prime}}_i^a{A^{\prime}}_i^a\right] \right.\right. \\[2mm] 
\left.\left. {\times }{\Pi }_{n=1}^N\displaystyle 
exp\left[i{\int }_{P_n}^{Q_n}{\sum}_{i=1}^2
A^{\prime(n)}_i(x)dx^i\right]{\Psi }_0\right\}\right)[{\Phi}] ~,\\[2mm]
\Phi =(A_1,B,c,\bar{c}), ~~ {\Xi}=(A_{\mu},B,c,\bar{c}), ~{\mu}=0,1,2~~~.
\label{pathform}
\end{array}
\end{equation}
where $\Psi _0$ is the vacuum state functional at time $t=-{\infty }$ and 
it is determined by Eq.(\ref{gausslaw}). Eq.(\ref{pathform}) is in fact the
gauge-fixed version of the state functional given in ref.\cite{lr}.
Therefore, we have that 
\begin{equation}
\begin{array}{l}
\hat{Q}|\text{phys}{\rangle }=\left\{ \displaystyle{\int }d^2x
~\left[ \frac k {8\pi }{\epsilon }^{ij}D_i\hat{c}^aA_j^a
-\frac 14f^{abc}\hat{\dot{\bar{c}}^a}\hat{c}^b
\hat{c}^c-\frac{ik}{8\pi }\hat{B}^aD^0\hat{c}^a\right] \right. \\[2mm] 
\left. \displaystyle {\times }{\Pi }_{n=1}^Nexp\left[i{\int }_{P_n~({\Gamma }%
_n)}^{Q_n}{\sum }_{i=1,2}\hat{A}_i^{(n)}(x_1,x_2)\right]\right\}|0{\rangle } \\%
[2mm] 
=\displaystyle \left\{ {\int }d^2x\left[-\hat{c}^a[\frac k{8\pi }\hat{F}
_{12}^a-{\sum }_{n=1}^NT_{(n)}^a({\delta }^{(2)}({\bf x}-{\bf x}_{P_n})-{%
\delta }^{(2)}({\bf x}-{\bf x}_{Q_n})]\right. \right. \\[2mm] 
\left. \left. +\displaystyle \frac k{4\pi }\hat{B}^a\hat{\Pi}_c^a-\frac 12%
f^{abc}\hat{\Pi}_c^a\hat{c}^b\hat{c}^c\right] \right\} |\text{phys}{\rangle }
\\[2mm] 
=\displaystyle \left\{ {\int }d^2x\left[-\hat{c}^a[\frac k{8\pi }\hat{F}%
_{12}^a-{\sum }_{n=1}^NT_{(n)}^a({\delta }^{(2)}({\bf x}-{\bf x}_{P_n})-{%
\delta }^{(2)}({\bf x}-{\bf x}_{Q_n})]\right] \right\} |\text{phys}{\rangle }
\\[2mm] 
{\oplus}|{\chi }_1\rangle >{\oplus}|{\chi }_2\rangle >=0~,
\end{array}
\end{equation}
where Eq.(\ref{brstcharge}) and the following operator 
equations have been used: 
\begin{equation}
\begin{array}{l}
\displaystyle\hat{A}_2^a=\frac{8\pi}{k}\hat{\Pi}^a_{A_1}=-\frac{8i\pi}{k}
\frac{\delta}{{\delta}\hat{A}_1^a}~~,\\[2mm]
\displaystyle [~{\partial }_{P_n}^{{\bf x}}\frac \delta {{\delta }\hat{A}_1^a({\bf x}_{P_n})}~,~
exp\left[i{\int }_{P_n~({\Gamma }_n)}^{Q_n}{\sum }_{i=1,2}\hat{A}_i^{(n)}({\bf x})dx^i\right]~]
=-i~T_{(n)}^a{\delta }^{(2)}(
{\bf x}-{\bf x}_{P_n})\\[2mm]
{\times}\displaystyle exp\left[i{\int }_{P_n~({\Gamma }_n)}^{Q_n}{\sum }_{i=1,2}\hat{A}_i^{(n)}({\bf x})dx^i\right]~~, \\[2mm] 
\displaystyle [~{\partial }_{Q_n}^{{\bf x}}
\frac \delta {{\delta }\hat{A}_1^a({\bf x}_{Q_n})}~,~
exp\left[i{\int }_{P_n~({\Gamma }_n)}^{Q_n}{\sum }_{i=1,2}\hat{A}_i^{(n)}({\bf x})dx^i\right]~]
=i~T_{(n)}^a{\delta }^{(2)}({\bf x}-{\bf x}_{Q_n})\\[2mm]
\displaystyle {\times} exp\left[i{\int }_{P_n~({\Gamma }_n)}^{Q_n}{\sum }_{i=1,2}\hat{A}_i^{(n)}({\bf x})dx^i\right]~~.
\end{array}
\end{equation}
Thus, the physical state condition $\hat{Q}|\text{phys}{\rangle }=0$, can be
reduced to the form 
\begin{equation}
\left[ \frac k{8\pi }\hat{F}_{12}^a-{\sum }_{n=1}^NT_{(n)}^a({\delta }^{(2)}(%
{\bf x}-{\bf x}_{P_n})-{\delta }^{(2)}({\bf x}-{\bf x}_{Q_n}))\right] |\text{%
phys}{\rangle }=0~.  \label{gaussl}
\end{equation}
These are exactly the Gauss law constraints given by Witten$^{\cite{wit}}$
in the case that Wilson loop operators are present.

{\large {\bf Acknowledgment}}: WFC is grateful to the World
Laboratory, Switzerland, for financial support.

\end{document}